\documentclass[11pt]{amsart}
\usepackage{graphicx}
\usepackage{amsfonts}
\usepackage{amsmath}
\usepackage{amssymb}
\usepackage{amsthm}
\usepackage{mathtools}
\usepackage{fullpage}
\usepackage{lmodern}
\usepackage{xcolor}
\usepackage{enumitem}
\usepackage{hyperref}
\hypersetup{
    colorlinks=true,
    linkcolor=purple,
    citecolor=brown,
    filecolor=red,
    urlcolor=blue,
    }
\usepackage[boxsize=5pt]{ytableau}
\usepackage{mathrsfs}
\allowdisplaybreaks

\newcommand{\CT}{\mathcal{T}}

\newcommand{\scT}{\mathscr{T}}

\newcommand{\CE}{\mathcal{E}}

\def\hgt{\mathrm{ht}}

\newcommand{\SP}{\mathsf{P}}

\newcommand{\SQ}{\mathsf{Q}}
\newcommand{\sQ}{\SQ}
\newcommand{\sU}{\mathsf{U}}

\newcommand{\IZ}{\mathbb{Z}}

\newcommand{\IC}{\mathbb{C}}

\newcommand{\Sk}{\mathrm{Sk}}
\newcommand{\SkAlg}{\mathrm{SkAlg}}

\newcommand{\sZ}{\mathsf{Z}}
\newcommand{\sP}{\mathsf{P}}
\newcommand{\sW}{\mathsf{W}}
\newcommand{\sPsi}{\mathsf{\Psi}}
\newcommand{\sPhi}{\mathsf{\Phi}}

\newcommand{\be}{\begin{equation}}
\newcommand{\ee}{\end{equation}}

\theoremstyle{plain}
\newtheorem{theorem}{Theorem}[section]

\newtheorem{corollary}[theorem]{Corollary}

\theoremstyle{definition}

\theoremstyle{remark}

\title{The topological vertex is finitely generated}
\author{Kunal Gupta}
\address{Department of Physics and Astronomy, Uppsala University, Box 516, 751 20 Uppsala, Sweden}
\email{kunal.gupta@physics.uu.se}
\author{Pietro Longhi}
\address{Department of Physics and Astronomy, Uppsala University, Box 516, 751 20 Uppsala, Sweden\newline
Department of Mathematics, Uppsala University, Box 480, 751 06 Uppsala, Sweden\newline
Centre for Geometry and Physics, Uppsala University, Box 516, 751 20 Uppsala, Sweden}
\email{pietro.longhi@physics.uu.se}

\begin{document}

\maketitle

\begin{abstract}
The topological vertex computes all genus amplitudes for the open topological A-model in $\mathbb{C}^3$ with branes on three toric Lagrangians and, via gluing, both open and closed string amplitudes on any toric Calabi-Yau threefold.
In this note we show that the topological vertex is generated by boundary interactions of three disks and three annuli. A generalization of this result is shown to hold for the refined topological vertex.
\end{abstract}

%
%\tableofcontents

%\newpage

\section{Introduction}

The topological vertex of \cite{Aganagic:2003db} counts open holomorphic curves in $\mathbb{C}^3$ with boundaries on three toric Lagrangians $L_i\simeq S^1\times D^2$.
Topological gauge/string duality identifies the open string partition function with the path integral of $U(N_i)$ Chern-Simons gauge theory on each $L_i$ \cite{Witten:1992fb}. 
This in turn computes a state in the Hilbert space associated to the boundary torus $\partial L_i\simeq T^2$, a model for which is provided by the HOMFLYPT skein module of $\Sk_{q,a_i}(S^1\times D^2)$ with the identification $a_i=q^{N_i/2}$ \cite{Witten:1988hf}. This interplay between enumerative geometry and TQFT has been placed on a firm mathematical footing by the recent development of skein-valued curve counting \cite{SoB}. A geometric interpretation of the topological vertex was advanced in this setting by \cite{Ekholm:2024lir}, which identified appropriate asymptotic resolution of Lagrangians $L_i$ that gives rise to the result of \cite{Aganagic:2003db}.

\begin{figure}[h!]
\begin{center}
\includegraphics[width=0.32\textwidth]{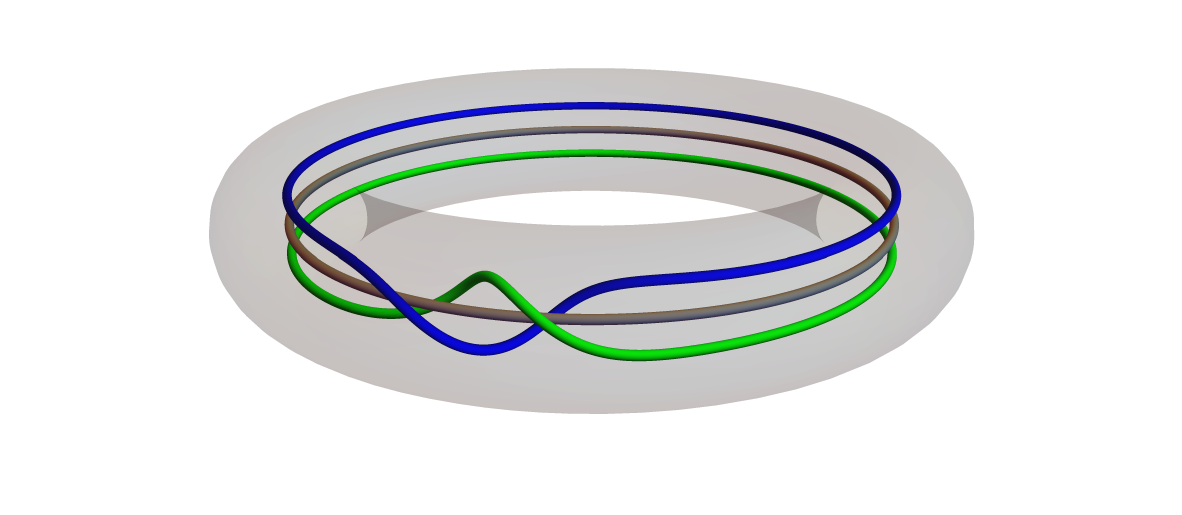}
\includegraphics[width=0.32\textwidth]{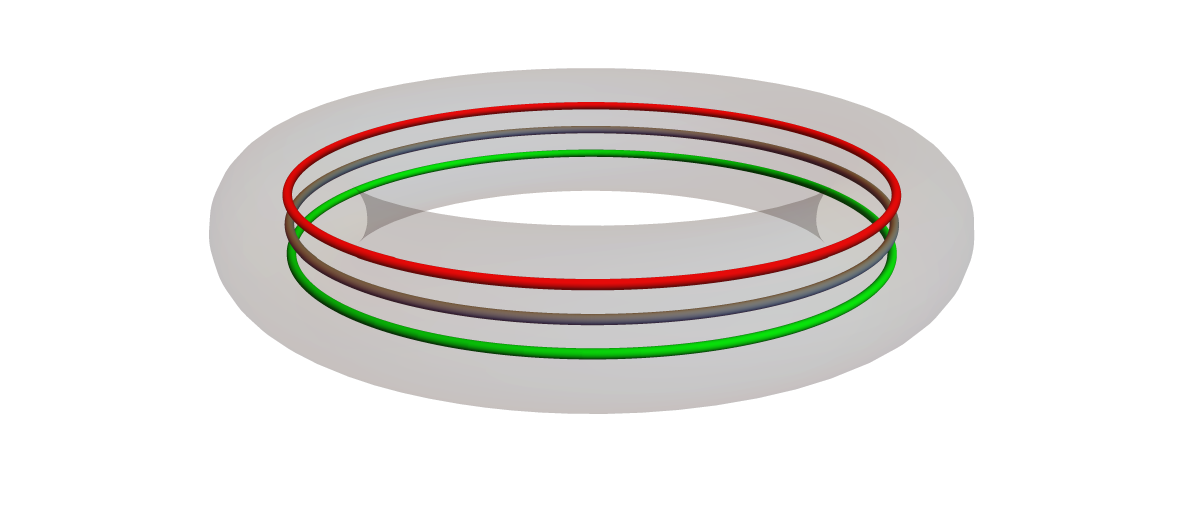}
\includegraphics[width=0.32\textwidth]{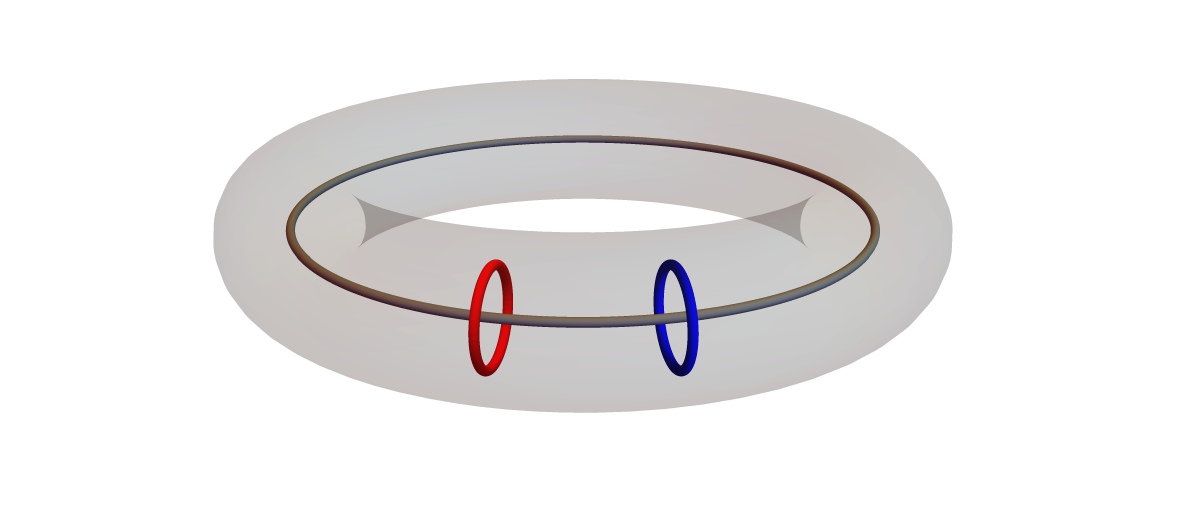}
\caption{Boundaries of basic curves on the three toric Lagrangians $L_1, L_2, L_3$.
Three disk boundaries are shown in gray. Anti-annuli in red, green and blue.}
\label{fig:curves}
\end{center}
\end{figure}

The first main result of this paper is a new formula for the open string partition function of three toric branes in $\IC^3$.  
We define operators $\sPsi_{\mathrm{di}}^{(\alpha|i,j)}\in \SkAlg(T^2_\alpha)$ and $\sPsi_{\mathrm{aa}}^{(\alpha|i,j)(\beta|k,l)}\in \SkAlg(T^2_\alpha)\otimes \SkAlg(T^2_\beta)$ labeled by $\alpha,\beta\in\{1,2,3\}$ and by $(i,j)\in H_1(T^2_\alpha)$ and $(k,l)\in H_1(T^2_\beta)$.
When applied to the empty skein element
$\sW_{\emptyset}\otimes\sW_{\emptyset}\otimes\sW_{\emptyset}\in \Sk(\bigsqcup_{i=1}^{3} L_i)$,
 these operators create framed disks and (anti)annuli with the specified boundary homology classes $(i,j)$ descending to winding $j\in H_1(L_\alpha,\IZ)$ and framing $i$. We show that the topological vertex partition function is generated by insertions of unframed disks on each $L_i$ and three anti-annuli stretched between each pair of branes
\be\label{eq:intro-main-result}
	\begin{split}
	\sZ
	= 
	\;&
	\sPsi_{\mathrm{aa}}^{(1|-1, 1)(2|0, 1)}  
	\cdot
	\sPsi_{\mathrm{aa}}^{(2|0, 1)(3|1, 0)}[a_{L_3}^{-1}]  
	\cdot
	\sPsi_{\mathrm{aa}}^{(3|-1, 0)(1|-1, 1)}[a_{L_3}]
	\\
	\cdot &\,
	\sPsi_{\mathrm{di}}^{(3|0,1)}
	\cdot
	\sPsi_{\mathrm{di}}^{(2|0,1)}[a_{L_3}^{-2}] 
	\cdot 
	\sPsi_{\mathrm{di}}^{(1|0,1)}[a_{L_3}^2]
%	\\
	\cdot 
%	&\,
	\sW_{\emptyset}\otimes\sW_{\emptyset}\otimes\sW_{\emptyset}\,.
\end{split}
\ee
The three disk operators create disks in each $L_i$ when acting on the empty skein, these are then linked by boundaries of anti-annuli as shown in Figure \ref{fig:curves}. 
Additional details on this formula, and a generalization to arbitrary framing of the three branes are provided in Section~\ref{sec:unrefined-vertex}.

The second main result is an analogue of \eqref{eq:intro-main-result} for the refined topological vertex, whose explicit form is given in \eqref{eq:refined-quiver}. A novelty is that $\IZ_3$ symmetry among $L_i$ is broken by the existence of a `preferred direction' in the definition of the refined topological vertex \cite{Iqbal:2007ii}. Nevertheless we find that the refined topological vertex is again finitely generated by three disk and three annuli operators. The definition of these operators takes values in the Elliptic Hall Algebra $\CE_{q,t}$ of \cite{burban,schiffmann2011elliptic} which is a deformation of the skein algebra of the torus \cite{morton}. 
We prove that the refined topological vertex is finitely generated for a specific choice of framing for each of the three branes. 
\medskip

\noindent
The results of this note raise a number of questions that we leave to future work
\begin{enumerate}
\item
It would be important to clarify the geometric meaning of formula \eqref{eq:intro-main-result}.
In particular the anti-annuli with boundaries on $L_3$ do not seem to arise as actual holomorphic curves in the geometric setting. For this reason our formula should be regarded as a combinatorial statement rather than a geometric one at this stage.
\item
Finiteness of the topological vertex suggests that open and closed topological string all genus amplitudes on any toric Calabi-Yau threefold can be formulated in terms of finitely many disks and annuli, with appropriate gluing data. It would be very interesting to explore the mechanisms that give rise to this, and a natural starting point would be to review the recent results of \cite{hu2025skein} in light of \eqref{eq:intro-main-result}. 
\item
Another interesting question concerns novel identities involving disks and annuli with different linking configurations. In \cite{scharitzer2023skein, nakamura2026recursion} a skein uplift of the pentagon identity for quantum dilogarithms was shown to hold. The identity relates a configuration of two disks with different units of linking, where a boundstate disk is created in the process of unlinking. Novel identities relating curves with different linking configurations are suggested by formula \eqref{eq:intro-main-result}, with some important differences.
The $\IZ_3$ symmetry of the topological vertex implies that the disks and the annulus with boundaries on $(L_\alpha, L_{\alpha+1})$ (indices understood mod 3) should generate a 2-brane partition function which is identical to that of disks and annuli on the pair of branes  $(L_{\alpha+1}, L_{\alpha+2})$  and on $(L_{\alpha+2}, L_{\alpha})$. However the linking configurations of disks and annuli involved for each pair of branes differ, making this a nontrivial identity. Moreover since this identity involves disks and annuli, as opposed to only disks, it is unclear whether earlier results imply its validity. See \cite[Appendix A]{Ekholm:2019lmb} for a related discussion in the context of $U(1)$ skein modules. The second important difference is that our identities hold in the skein of the solid torus, and it remains an open question whether they lift to the skein algebra.
\end{enumerate}

\section{Basic curves for the topological vertex}\label{sec:unrefined-vertex}

The open topological string partition function of three noncompact toric Lagrangians $L_1,L_2, L_3$ in $\IC^3$ is 
\be\label{eq:vertex-partition-function}
	\sZ
	:= \sum_{\lambda_1,\lambda_2,\lambda_3} \CT_{\lambda_1,\lambda_2,\lambda_3}(q) \, \sW_{\lambda_1}\otimes \sW_{\lambda_2}\otimes \sW_{\lambda_3}
\ee
where
\be\label{eq:topological-vertex}
	\CT_{\lambda_1,\lambda_2,\lambda_3}(q) 
	:= 
	(-1)^{|\lambda_1|+|\lambda_2|+|\lambda_3|}
	q^{-\kappa(\lambda_1)/2-\kappa(\lambda_3)/2} 
	\, s_{\lambda_3}(q^{\rho}) \, \sum_\eta s_{\lambda_1/\eta}(q^{\lambda_3^t + \rho}) s_{\lambda_2^t/\eta}(q^{\lambda_3 + \rho}),
\ee
is the topological vertex of \cite{Aganagic:2003db} expressed with conventions of \cite{Ekholm:2024lir} in terms of Schur functions $s_\mu$, skew-Schur functions $s_{\mu/\eta}$, and their principal specialization defined by $\rho = (-\frac12, -\frac32, \ldots)$, with the definition $ \kappa(\lambda) = \sum_{i} (\lambda_i)^2 -  \sum_{j} (\lambda^{t}_j)^2$.
The partition function \eqref{eq:vertex-partition-function} takes values in the HOMFLYPT skein $\bigotimes_{i=1,2,3}\Sk_+(L_i)$ \cite{SoB}, which realizes the space of Wilson loops $\sW_{\lambda}$ in the worldvolume Chern-Simons theory on the branes \cite{Witten:1988hf, Witten:1992fb}, and 
counts holomorphic curves with boundaries on $L_1\sqcup L_2\sqcup L_3$ with a resolution for the Legendrian boundaries $\partial L_i\subset  \partial \IC^3$ given by the Reeb flow at infinity \cite{Ekholm:2024lir}. We also have the corresponding framed objects
\begin{equation}
    \begin{split}
        \CT_{\lambda_1,\lambda_2,\lambda_3}^{f_1,f_2,f_3}(q) &:= \prod_{i=1}^{3} \left((-1)^{|\lambda_i|} q^{\kappa(\lambda_i)/2}\right)^{f_i} \CT_{\lambda_1,\lambda_2,\lambda_3}(q)\,,
        \\ 
        \sZ_{f_1, f_2, f_3}
	&:=\sum_{\lambda_1,\lambda_2,\lambda_3} \CT^{f_1, f_2, f_3}_{\lambda_1,\lambda_2,\lambda_3}(q) \, \sW_{\lambda_1}\otimes \sW_{\lambda_2}\otimes \sW_{\lambda_3}\,.
    \end{split}
\end{equation}

\begin{theorem}\label{thm:unrefined-quiver}
The skein-valued partition function \eqref{eq:vertex-partition-function}  is generated by three disks and three anti-annuli.
\be\label{eq:unrefined-quiver}
\begin{split}
	\sZ
	= 
	\;&
	\sPsi_{\mathrm{aa}}^{(1|-1, 1)(2|0, 1)}  
	\cdot
	\sPsi_{\mathrm{aa}}^{(2|0, 1)(3|1, 0)}[a_{L_3}^{-1}]  
	\cdot
	\sPsi_{\mathrm{aa}}^{(3|-1, 0)(1|-1, 1)}[a_{L_3}]
	\\
	\cdot &\,
	\sPsi_{\mathrm{di}}^{(3|0,1)}
	\cdot
	\sPsi_{\mathrm{di}}^{(2|0,1)}[a_{L_3}^{-2}] 
	\cdot 
	\sPsi_{\mathrm{di}}^{(1|0,1)}[a_{L_3}^2]
	\\
	\cdot &\,
	\sW_{\emptyset}\otimes\sW_{\emptyset}\otimes\sW_{\emptyset}\,.
\end{split}
\ee
\end{theorem}
To introduce the definition of the building blocks in expression \eqref{eq:unrefined-quiver} 
recall that the skein algebra of the torus $\Sk(T^2)$ admits a set of generators $\sP_{i,j}$ obeying
\be
	[\sP_{i,j} ,\sP_{k,l}] = (q^{(i,j)\wedge (k,l)/2} - q^{(k,l)\wedge (i,j)/2}) \sP_{i+k,j+l}\,.
\ee
Recall that the positive half of the algebra, corresponding to operators $\sP_{m,n}$ with $(m,n)\in \IZ\times \IZ_{\geq 0}$ acting as follows on $\Sk_+(S^1\times D^2)$, the part of the skein corresponding to positive winding around $H_1(S^1\times D^2)$ \cite{morton} 
\be\label{eq:generalformulaePW}
\begin{split}
	\sP_{0,0} \, \sW_{\lambda} 
	& = 
	\frac{a_L-a_L^{-1}}{q^{1/2}-q^{-1/2}} \,  \sW_{\lambda} \,,
	\\
	\sP_{m,0} \, \sW_{\lambda} 
	& = 
	\left(\frac{a_L^m-a_L^{-m}}{q^{m/2}-q^{-{m/2}}} 
	+ (q^{m/2}-q^{-{m/2}}) 
	a_L^m 
    % C_\lambda(q^{m})
    \sum_{(i,j)\in\lambda}q^{m(j-i)}
	\right)\, \sW_{\lambda} \,,
	\\
	\sP_{0,n} \, \sW_{\lambda} 
	& = 
	\sum_{\alpha \in \lambda+[n]} (-1)^{\hgt(\alpha-\lambda)}\sW_{\alpha}\,,
\end{split}
\ee
where $a_L$ denotes the HOMFLYPT skein parameter related to the number of branes on $L$, which counts intersections of holomorphic curves with the 4-chain of $L$ \cite{SoB, Ekholm:2024ceb}.
Using these we define the skein-algebra valued operators (for $(i,j), (k,l)\neq (0,0)$)
\be\label{eq:operator-partition-functions}
\begin{split}
	\sPsi_{\mathrm{di}}^{(\alpha| i,j)}[\xi] &
	:= \exp\left(
	-\sum_{d\geq 1}\frac{1}{d} \, \frac{(-a_{L_\alpha})^{-d i j}}{q^{d/2}-q^{-d/2}} \,\xi^d\, \sP^{(\alpha)}_{di,dj}
	\right),
	\\
	\sPsi_{\mathrm{aa}}^{(\alpha|i,j)(\beta| k,l)}[\xi] &
	:= \exp\left(
	-\sum_{d\geq 1} \frac{(-a_{L_\alpha})^{-d i j} (-a_{L_\beta})^{-d kl} }{d}  \,\xi^d\, \sP^{(\alpha)}_{di,dj}\otimes \sP^{(\beta)}_{dk,dl}
	\right).
\end{split}
\ee
These operators create framed disks and anti-annuli when acting on the empty skein 
\be\label{eq:skein-operators-on-empty-skein}
\begin{split}
	\sPsi_{\mathrm{di}}^{(\alpha|f,1)}[\xi]  \cdot \sW^{(\alpha)}_{\emptyset} 
	& = \sum_\lambda \left((-1)^{|\lambda|} q^{\kappa(\lambda)/2}\right)^{f} \left(\prod_{\ydiagram{1}\in\lambda} \frac{-q^{-c(\ydiagram{1})/2}}{q^{h(\ydiagram{1})/2}-q^{-h(\ydiagram{1})/2}}\right)  \,\xi^{|\lambda|}\, \sW^{(\alpha)}_{\lambda}\,,
	\\
	\sPsi_{\mathrm{aa}}^{(\alpha|f_1, 1)(\beta|f_2, 1)}[\xi]  \cdot\sW^{(\alpha)}_{\emptyset} \otimes \sW^{(\beta)}_{\emptyset} 
	& = \sum_{\lambda}
	\left((-1)^{|\lambda|} q^{\kappa(\lambda)/2}\right)^{f_1-f_2}
	\,\xi^{|\lambda|}\, 
	(-1)^{|\lambda|} \sW^{(\alpha)}_{\lambda} \otimes \sW^{(\beta)}_{\lambda^t} \,.
\end{split}
\ee
Recall from \cite[Eqs.~(5.4),(5.5),(5.6)]{Ekholm:2024ceb} that a change of framing acts on $\Sk(S^1\times D^2)$ and on $\SkAlg(T^2)$ as follows
\begin{equation}\label{eq:framing-change}
    \sW_\lambda \to (-1)^{|\lambda|}q^{\kappa(\lambda)/2}\sW_\lambda, \qquad \SP_{i,j} \to (-a_L)^{-j}\SP_{i+j,j},
\end{equation}
which using Theorem~\ref{thm:unrefined-quiver} implies the following corollary.
\begin{corollary}
The framed skein-valued partition function admits the following expression
\be
\begin{split}
	\sZ_{f_1, f_2, f_3}
	=\;&
	\sPsi_{\mathrm{aa}}^{(1|f_1-1, 1)(2|f_2, 1)}  
	\cdot
	\sPsi_{\mathrm{aa}}^{(2|f_2, 1)(3|1, 0)}[a_{L_3}^{-1}]  
	\cdot
	\sPsi_{\mathrm{aa}}^{(3|-1, 0)(1|f_1-1, 1)}[a_{L_3}]
	\\
	\cdot &\,
	\sPsi_{\mathrm{di}}^{(3|f_3,1)}
	\cdot
	\sPsi_{\mathrm{di}}^{(2|f_2,1)}[a_{L_3}^{-2}] 
	\cdot 
	\sPsi_{\mathrm{di}}^{(1|f_1,1)}[a_{L_3}^2]
	\\
	\cdot &\,
	\sW_{\emptyset}\otimes\sW_{\emptyset}\otimes\sW_{\emptyset}\,.
\end{split}	
\ee
\end{corollary}

\section{Generalization for the refined vertex}\label{sec:refined-vertex}
The refined topological vertex introduced in \cite{Iqbal:2007ii}, see also \cite{Awata:2008ed, Awata:2011ce} is
\be\label{eq:refined-vertex-def}
\begin{split}
    \scT_{\lambda_1,\lambda_2,\lambda_3}(t,q)
    := {}& q^{\|\lambda_1^t\|^2/2}t^{-\|\lambda_1\|^2/2}
    \left(\frac qt\right)^{|\lambda_3|/2}
    P_{\lambda_3}(t^{-\rho};q,t)\sum_\eta\left(\frac qt\right)^{|\eta|/2}
    s_{\lambda_1^t/\eta}(t^{-\rho}q^{-\lambda_3})
    s_{\lambda_2/\eta}(q^{-\rho}t^{-\lambda_3^t})\,,
\end{split}
\ee
where $||\lambda||^2 := \sum_i |\lambda_i|^2$ and  $P_\lambda(X;q,t)$ are Macdonald functions \cite{macdonald1998symmetric}. 
The above has been suitably normalised so that we recover \eqref{eq:topological-vertex} in the limit $t=q$, see \eqref{eq:comparison-claim}.
We define the corresponding refined topological string partition function
\be\label{eq:refined-vertex-partition-function}
	 \sZ^{q,t} := \sum_{\lambda_1,\lambda_2,\lambda_3} \scT_{\lambda_1,\lambda_2,\lambda_3}(t,q)\, \sW_{\lambda_1}\otimes \sW_{\lambda_2}\otimes \sQ_{\lambda_3}\,,
\ee
where $\sQ_\lambda$ is a basis for $\Sk_+(S^1\times D^2)$ related to $\sW_\lambda$ 
by the degree-preserving upper-triangular change of basis between Schur polynomials and dual Macdonald functions, see \eqref{eq:skein-Lambda-iso}.
 The corresponding framed objects are defined as
\begin{equation}\label{eq:refined-vertex-partition-function-framed}
    \begin{split}
        \scT_{\lambda_1,\lambda_2,\lambda_3}^{f_1,f_2,f_3}(t,q) &:= \prod_{i=1}^3((-1)^{|\lambda_i|}t^{n(\lambda_i^t)}q^{-n(\lambda_i)})^{f_i} \scT_{\lambda_1,\lambda_2,\lambda_3}(t,q)\,, \\
         \sZ^{q,t}_{f_1,f_2,f_3} &:=  \sum_{\lambda_1,\lambda_2,\lambda_3} \scT_{\lambda_1,\lambda_2,\lambda_3}^{f_1,f_2,f_3}(t,q)\, \sW_{\lambda_1}\otimes \sW_{\lambda_2}\otimes \sQ_{\lambda_3}\,,
    \end{split}
\end{equation}
where $n(\lambda) := \sum_i(i-1)\lambda_i=\frac{1}{2} ( ||\lambda^t||^2 - |\lambda|)$. 
We introduce a deformation of the skein algebra of $T^2$ generated by $\sU_{m,n}$, where for $(m,n) \in \mathbb{Z}\times \mathbb{Z}_{\geq 0}$ the action on $\Sk_+(S^1\times D^2)$ is defined as follows 
\be\label{eq:EHA-action}
\begin{split}
    \sU_{0,0} \sQ_\lambda &= \frac{a_L-a_L^{-1}}{t^{1/2}-t^{-1/2}} \,\sQ_\lambda\,,\\
    \sU_{m,0}\sQ_\lambda&= \left(\frac{a_L^m-a_L^{-m}}{t^{m/2}-t^{-m/2}}+ a_L^{m}(q^{m/2}-q^{-m/2})\sum_{(i,j)\in\lambda}q^{m(j-1/2)}t^{-m(i-1/2)}\right)\sQ_\lambda\,,\\
    \sU_{0,n}\sQ_\lambda
    &= \SP_{0,n}\,\sQ_\lambda\,,
\end{split}
\ee
It is shown in \cite{Gupta-Longhi-EHA-recursion} that after extending the action of these operators to the full $\Sk(S^1\times D^2)$ and including additional generators $\sU_{0,-n}$, these provide a representation of the Elliptic Hall algebra of \cite{burban, schiffmann2011elliptic}.
In the limit $t=q$ these definitions recover the Schur basis of $\Sk_+(S^1\times D^2)$ and the positive half of $\SkAlg(T^2)$,
\be\label{eq:unrefined-limit}
    \lim_{t\to q}\sQ_\lambda = \sW_\lambda,
    \qquad
    \lim_{t\to q}\sU_{0,n} = \sP_{0,n},
    \qquad
    \lim_{t\to q}\sU_{m,0} = \sP_{m,0},
    \qquad (m\in \IZ, \ n\geq1).
\ee
Using these, we define refined operator-valued partition functions for disk and anti-annulus
\begin{align}
    \sPhi_{\mathrm{di}}^{(\alpha|i,j)}[\xi]
    &= \exp\left(-\sum_{d\geq1}\frac{\xi^d}{d(q_\alpha^{d/2}-q_\alpha^{-d/2})}\left(\frac{q_\alpha}{t_\alpha}\right)^{d/2}\sU^{(\alpha)}_{di,dj}\right),
    \label{eq:refined-disk-operator}\\
    \sPhi_{\mathrm{aa}}^{(\alpha|i,j)(\beta|k,l)}[\xi]
    &= \exp\left(-\sum_{d\geq1}\frac{\xi^d}{d}\frac{1-t_\alpha^d}{1-q_\beta^d}\sU^{(\alpha)}_{di,dj}\otimes\sU^{(\beta)}_{dk,dl}\right).
    \label{eq:refined-annulus-operator}
\end{align}
where $(i,j),(k,l) \in \{(\pm 1,0),(0,1)\}$ and $(q_1,t_1)=(t,q)$, $(q_2,t_2)=(q_3,t_3)=(q,t)$. 
Their action on the empty skein is given in \eqref{eq:refined-disk} and \eqref{eq:refined-annulus-1}.

\begin{theorem}\label{thm:refined-quiver}
The skein-valued partition function of the refined vertex \eqref{eq:refined-vertex-partition-function-framed} 
in framing $(1,0,0)$ admits the following expression  
\be\label{eq:refined-quiver}
\begin{split}
     \sZ^{q,t}_{1,0,0}
    =\;&
    \sPhi_{\mathrm{aa}}^{(1|0,1)(2|0,1)}\left[\frac qt\right]
    \sPhi_{\mathrm{aa}}^{(2|0,1)(3|1,0)}\left[a_{L_3}^{-1}\sqrt{\frac qt}\right]\cdot
    \sPhi_{\mathrm{aa}}^{(3|-1,0)(1|0,1)}\left[a_{L_3}\sqrt{\frac qt}\right]\\
    &\cdot
    \left(\sPhi_{\mathrm{di}}^{(1|0,1)}\left[a_{L_3}^{2}\frac qt\right]\right)^{-1}
    \sPhi_{\mathrm{di}}^{(2|0,1)}\left[a_{L_3}^{-2}\sqrt{\frac tq}\right]
    \sPhi_{\mathrm{di}}^{(3|0,1)}\left[\sqrt{\frac tq}\right]
    \\&\cdot\sW_{\emptyset}\otimes\sW_{\emptyset}\otimes\sW_{\emptyset}\,.
\end{split}
\ee
\end{theorem}

\section{Proof of Theorems \ref{thm:unrefined-quiver} and \ref{thm:refined-quiver}}
Recall the definition of Cauchy kernel for two alphabets $X$ and $Y$
    \begin{equation}\label{eq:Cauchy-Kernel}
        \Omega(X,Y) := \exp\left(\sum_{d\geq1}\frac{p_d(X)p_d(Y)}{d}\right),
    \end{equation}
    and the generalized Cauchy identities, see e.g. \cite[Chapters 1, 6]{macdonald1998symmetric}
    \begin{align}
    \sum_\lambda s_{\lambda/\eta}(X)s_\lambda(Y)
    &= \Omega(X,Y)s_\eta(Y)\,,
    \label{eq:Cauchy-Identites-1}\\
    \sum_\lambda (-c)^{|\lambda|}s_{\lambda^t/\eta}(X)s_\lambda(Y)
    &= \frac{(-c)^{|\eta|}s_{\eta^t}(Y)}{\Omega(cX,Y)}\,,
    \label{eq:Cauchy-Identites-2}\\
    \sum_\lambda P_\lambda(X;q,t)Q_\lambda(Y;q,t)
    &= \exp\left(\sum_{d\geq1}\frac{p_d(q^{-\rho-1/2})}{p_d(t^{-\rho-1/2})}\frac{p_d(X)p_d(Y)}{d}\right)\,.
    \label{eq:Cauchy-Identites-3}
    \end{align}
    
\begin{proof}[Proof of Theorem~\ref{thm:refined-quiver}]
Using the isomorphism between $\Sk_+(S^1\times D^2)$ and the ring of symmetric functions \cite{morton} we identify
    \begin{equation}\label{eq:skein-Lambda-iso}
        \sW_{\lambda_1}\otimes\sW_{\lambda_2}\otimes\sQ_{\lambda_3} =  s_{\lambda_1}(X) s_{\lambda_2}(Y)  Q_{\lambda_3}(Z;q,t)\,,
    \end{equation}
    where $X,Y,Z$ denote auxiliary alphabets for the first, second and third leg of the refined vertex, and the dual Macdonald functions $Q_\lambda(Z;q,t)$ are related to Schur polynomials $s_{\lambda}(Z)$ by a degree-preserving upper-triangular change of basis, see \cite[Chapter VI, Section 4]{macdonald1998symmetric}.
    The left hand side of \eqref{eq:refined-quiver} then becomes
    \begin{equation}
    \begin{split}
     \sZ^{q,t}_{1,0,0}
    = \sum_{\lambda_1,\lambda_2,\lambda_3,\eta}&(-1)^{|\lambda_1|}
    \left(\frac qt\right)^{(|\lambda_1|+|\lambda_3|+|\eta|)/2}
    P_{\lambda_3}(t^{-\rho};q,t)\,
    s_{\lambda_1^t/\eta}(t^{-\rho}q^{-\lambda_3})s_{\lambda_2/\eta}(q^{-\rho}t^{-\lambda_3^t})\,\\&
    \times s_{\lambda_1}(X)s_{\lambda_2}(Y)Q_{\lambda_3}(Z;q,t)\,.
    \end{split}
    \end{equation}
    Sums over $\lambda_1$ and $\lambda_2$ can be performed by using respectively \eqref{eq:Cauchy-Identites-2} and  \eqref{eq:Cauchy-Identites-1}, which gives
    \begin{equation}
     \sZ^{q,t}_{1,0,0}
    = \sum_{\lambda_3,\eta}\left(\frac qt\right)^{|\lambda_3|/2}
    \left(-\frac qt\right)^{|\eta|}
    P_{\lambda_3}(t^{-\rho};q,t)Q_{\lambda_3}(Z;q,t)\,
    \frac{\Omega(q^{-\rho}t^{-\lambda_3^t},Y)}{\Omega(\sqrt{q/t}\,t^{-\rho}q^{-\lambda_3},X)}
    s_{\eta^t}(X)s_\eta(Y)\,.
    \end{equation}
    Further using  \eqref{eq:Cauchy-Identites-2} for the summation over $\eta$, we get
    \begin{equation}\label{eq:refined-identity-lhs-explicit}
         \sZ^{q,t}_{1,0,0}
    = \frac{1}{\Omega((q/t)X,Y)}
    \sum_{\lambda_3}\left(\frac qt\right)^{|\lambda_3|/2}
    P_{\lambda_3}(t^{-\rho};q,t)Q_{\lambda_3}(Z;q,t)\,
    \frac{\Omega(q^{-\rho}t^{-\lambda_3^t},Y)}{\Omega(\sqrt{q/t}\,t^{-\rho}q^{-\lambda_3},X)}\,.
    \end{equation}
    Next we consider the right hand side of  \eqref{eq:refined-quiver}. 
    Note that the three disk operators mutually commute and the same is true for the three anti-annuli operators, although disk and anti-annuli operators do not commute with each other in general.
    The skein-valued operators corresponding to the refined disk and anti-annulus \eqref{eq:refined-disk-operator} and \eqref{eq:refined-annulus-operator} with $(i,j)=(k,l)=(0,1)$ act as follows on any basis element of the skein 
\begin{align}
    &\sPhi_{\mathrm{di}}^{(\alpha|0,1)}[\xi]\cdot\sW_\nu^{(\alpha)}
    = \sum_\lambda \xi^{|\lambda|}
    \left(\frac{q_\alpha}{t_\alpha}\right)^{|\lambda|/2}
    s_\lambda(q_\alpha^{-\rho})\,s_\lambda(X_{\alpha}) \, s_{\nu}(X_\alpha)
    =
    \Omega(\xi \sqrt{q_\alpha/t_\alpha} q_\alpha^{-\rho},X_\alpha)
    \, s_{\nu}(X_\alpha),
    \label{eq:refined-disk}\\
    &\sPhi_{\mathrm{aa}}^{(\alpha|0,1)(\beta|0,1)}[\xi]
    \cdot\left(\sW_\mu^{(\alpha)}\otimes\sW_\nu^{(\beta)}\right)
    = 
    \exp\left(-\sum_{d\geq 1}
    \frac{1}{d}\xi^d\frac{1-t_\alpha^d}{1-q_\beta^d} p_d(X_\alpha)p_d(X_\beta)\right)
    \, s_{\mu}(X_\alpha)\, s_{\nu}(X_\beta)\,,
    \label{eq:refined-annulus-1}
\end{align}
where $X_\alpha$ specializes to $X,Y,Z$ for $\alpha=1,2,3$ respectively.
    Using  \eqref{eq:refined-disk}, we immediately get the action of the three refined disks as
    \begin{equation}
    \begin{split}
    &\left(\sPhi_{\mathrm{di}}^{(1|0,1)}\left[a_{L_3}^{2}\frac qt\right]\right)^{-1}
    \sPhi_{\mathrm{di}}^{(2|0,1)}\left[a_{L_3}^{-2}\sqrt{\frac tq}\right]
    \sPhi_{\mathrm{di}}^{(3|0,1)}\left[\sqrt{\frac tq}\right]\cdot 1 = \frac{\Omega[a_{L_3}^{-2}q^{-\rho},Y]}
    {\Omega(a_{L_3}^{2}\sqrt{q/t}\,t^{-\rho},X)}\,
    \Omega(q^{-\rho},Z)\,.
\end{split}
\end{equation}
Next we introduce the anti-annulus on the first two legs using \eqref{eq:refined-annulus-1}, and further employ \eqref{eq:Cauchy-Identites-3} to write the Cauchy kernel involving $Z$ in terms of Macdonald functions
\begin{equation}\label{eq:3d-1a}
    \begin{split}
        &\sPhi_{\mathrm{aa}}^{(1|0,1)(2|0,1)}\left[\frac qt\right]\left(\sPhi_{\mathrm{di}}^{(1|0,1)}\left[a_{L_3}^{2}\frac qt\right]\right)^{-1}
    \sPhi_{\mathrm{di}}^{(2|0,1)}\left[a_{L_3}^{-2}\sqrt{\frac tq}\right]
    \sPhi_{\mathrm{di}}^{(3|0,1)}\left[\sqrt{\frac tq}\right]\cdot 1 \\
    & = \frac{1}{\Omega((q/t)X,Y)}\frac{\Omega(a_{L_3}^{-2}q^{-\rho},Y)}
    {\Omega(a_{L_3}^{2}\sqrt{q/t}\,t^{-\rho},X)}\,
    \Omega(q^{-\rho},Z)
    \\
    & = \frac{1}{\Omega((q/t)X,Y)}\frac{\Omega(a_{L_3}^{-2}q^{-\rho},Y)}
    {\Omega(a_{L_3}^{2}\sqrt{q/t}\,t^{-\rho},X)}\,
    \sum_\nu\left(\frac qt\right)^{|\nu|/2}
    P_\nu(t^{-\rho};q,t)Q_\nu(Z;q,t)\,.
    \end{split}
\end{equation}
Using the action \eqref{eq:EHA-action} in definition \eqref{eq:refined-annulus-operator} and the identities 
\begin{equation}
\begin{aligned}
    \sU^{(3)}_{-d,0}Q_\nu(Z;q,t)
    &=
    \left(a_{L_3}^{-d}p_d(t^{-\rho}q^{-\nu})
    -a_{L_3}^{d}p_d(t^{-\rho})\right)Q_\nu(Z;q,t)\,,\\
    \sU^{(3)}_{d,0}Q_\nu(Z;q,t)
    &=
    \left(\frac tq\right)^{d/2}\frac{1-q^d}{1-t^d}
    \left(a_{L_3}^{-d}p_d(q^{-\rho})
    -a_{L_3}^{d}p_d(q^{-\rho}t^{-\nu^t})\right)Q_\nu(Z;q,t)\,,
\end{aligned}
\end{equation}
the action of remaining anti-annulus operators for an arbitrary function $g(X,Y)$ is
\begin{equation}
\begin{aligned}
    \sPhi_{\mathrm{aa}}^{(3|-1,0)(1|0,1)}
    \left[a_{L_3}\sqrt{\frac qt}\right]
    \cdot\bigl(g(X,Y)Q_\nu(Z;q,t)\bigr)
    &=
    \frac{\Omega(a_{L_3}^{2}\sqrt{q/t}\,t^{-\rho},X)}
    {\Omega(\sqrt{q/t}\,t^{-\rho}q^{-\nu},X)}
    g(X,Y)Q_\nu(Z;q,t)\,,
    \\
    \sPhi_{\mathrm{aa}}^{(2|0,1)(3|1,0)}
    \left[a_{L_3}^{-1}\sqrt{\frac qt}\right]
    \cdot\bigl(g(X,Y)Q_\nu(Z;q,t)\bigr)
    &=
    \frac{\Omega(q^{-\rho}t^{-\nu^t},Y)}
    {\Omega(a_{L_3}^{-2}q^{-\rho},Y)}
    g(X,Y)Q_\nu(Z;q,t)\,.
\end{aligned}
\end{equation}
Using the above action on  \eqref{eq:3d-1a} immediately gives \eqref{eq:refined-identity-lhs-explicit}.
\end{proof}

\begin{proof}[Proof of Theorem~\ref{thm:unrefined-quiver}]
    Replacing $t=q$ in \eqref{eq:refined-vertex-def}  
    and using the identities (see e.g. \cite[Proposition 4.1]{zhou2003curve})
$
	s_\lambda(q^{-\rho}) = q^{-\kappa(\lambda)/2}s_{\lambda^t}(q^{-\rho}) = (-1)^{|\lambda|}q^{-\kappa(\lambda)/2}s_\lambda(q^{\rho})
$ and $s_{\lambda^t/\eta}(q^{-\nu-\rho}) = (-1)^{|\lambda|+|\eta|}s_{\lambda/\eta^t}(q^{\nu^t+\rho})$
gives \eqref{eq:topological-vertex} 
\be\label{eq:comparison-claim}
	\lim_{t\to q} \scT_{\lambda_1,\lambda_2,\lambda_3}(t,q)=\CT_{\lambda_1,\lambda_2,\lambda_3}(q) \,.
\ee
    Since $\lim_{t\to q}t^{n(\lambda_i^t)}q^{-n(\lambda_i)} = q^{\kappa(\lambda_i)/2}$ it also follows that $\lim_{t\to q} \sZ^{q,t}_{f_1,f_2,f_3} = \sZ_{f_1,f_2,f_3} $.
    Moreover using \eqref{eq:unrefined-limit} it is clear that the refined disk and anti-annulus operators \eqref{eq:refined-disk-operator} and \eqref{eq:refined-annulus-operator} reduce to \eqref{eq:operator-partition-functions}.
    From this it follows that \eqref{eq:refined-quiver} specializes to the identity
    \begin{equation}
\begin{split}
    \sZ_{1,0,0} 
    =\;& \sPsi_{\mathrm{aa}}^{(1|0,1)(2|0,1)}
    \cdot\sPsi_{\mathrm{aa}}^{(2|0,1)(3|1,0)}[a_{L_3}^{-1}]
    \cdot\sPsi_{\mathrm{aa}}^{(3|-1,0)(1|0,1)}[a_{L_3}]\\
    &\cdot
    \left(\sPsi_{\mathrm{di}}^{(1|0,1)}[a_{L_3}^{2}]\right)^{-1}
    \cdot\sPsi_{\mathrm{di}}^{(2|0,1)}[a_{L_3}^{-2}]
    \cdot\sPsi_{\mathrm{di}}^{(3|0,1)}
    \cdot \sW_{\emptyset}\otimes \sW_{\emptyset} \otimes \sW_{\emptyset}\,.
\end{split}
\end{equation}
Next observe that the anti-disk on the first Lagrangian can be traded for a framed disk using \eqref{eq:skein-operators-on-empty-skein} together with \eqref{eq:Cauchy-Identites-2}
\begin{equation}
\begin{split}
    \left(\sPsi_{\mathrm{di}}^{(1|0,1)}[a_{L_3}^{2}]\right)^{-1}
    \cdot\sW_\emptyset^{(1)}
    &
    = \sum_\lambda(-a_{L_3}^{2})^{|\lambda|}
    s_{\lambda^t}[q^{-\rho}]\sW_\lambda^{(1)}\\
    &=\sum_\lambda(-a_{L_3}^{2})^{|\lambda|}
    q^{\kappa(\lambda)/2}s_{\lambda}[q^{-\rho}]\sW_\lambda^{(1)}= \sPsi_{\mathrm{di}}^{(1|1,1)}[a_{L_3}^{2}]
    \cdot\sW_\emptyset^{(1)}\,,
\end{split}
\end{equation}
    which allows us to write
    \begin{equation}
\begin{split}
    \sZ_{1,0,0}
    =\;& \sPsi_{\mathrm{aa}}^{(1|0,1)(2|0,1)}
    \sPsi_{\mathrm{aa}}^{(2|0,1)(3|1,0)}[a_{L_3}^{-1}]
    \sPsi_{\mathrm{aa}}^{(3|-1,0)(1|0,1)}[a_{L_3}]\\
    &\cdot
    \sPsi_{\mathrm{di}}^{(1|1,1)}[a_{L_3}^{2}]
    \sPsi_{\mathrm{di}}^{(2|0,1)}[a_{L_3}^{-2}]
    \sPsi_{\mathrm{di}}^{(3|0,1)}\\&\cdot \sW_{\emptyset}\otimes \sW_{\emptyset} \otimes \sW_{\emptyset}\,.
\end{split}
\end{equation}
Using  \eqref{eq:framing-change} to shift framing to $\sZ_{0,0,0}$ then immediately implies the result.
\end{proof}

% \newpage

\subsection*{Acknowledgements} 
We thank Tobias Ekholm and Vivek Shende for discussions and insightful comments on our results.
The work of KG is supported by the Knut and Alice Wallenberg Foundation grant KAW2021.0170, and by the Olle Engkvists Stiftelse Grant 2180108. The work of P.L. is supported
by the Knut and Alice Wallenberg Foundation, KAW2020.0307 Wallenberg Scholar and by the Swedish
Research Council, VR 2022-06593, Centre of Excellence in Geometry and Physics at Uppsala University.
Preprint number UUITP-25/26.

\bibliographystyle{plain}
\bibliography{bibliography.bib}

\end{document}